\documentclass[journal,comsoc]{IEEEtran}
\normalsize

\usepackage{amsmath,amsfonts}
\usepackage{algorithm}
\usepackage{algorithmic}  
\usepackage{amsmath}  
\usepackage{array}
\usepackage{textcomp}
\usepackage{url}
\usepackage{verbatim}
\usepackage{bm}
\usepackage{amsthm,amsmath,amssymb}
\usepackage{mathrsfs}
\usepackage{setspace}
\usepackage{mdwmath}
\usepackage{mdwtab}
\usepackage{eqparbox}
\usepackage{fixltx2e}
\usepackage{cite}
\usepackage{graphicx}
\usepackage{epstopdf}
\graphicspath{{figures/}}
\usepackage{stfloats}
\usepackage[caption=false,font=normalsize,labelfont=sf,textfont=sf]{subfig}
\usepackage{soul,color,xcolor}
\usepackage{setspace}
\usepackage{caption}
\usepackage{balance}
\usepackage{booktabs}

\title{Rotatable RIS Assisted Physical-Layer Multicasting}
\author{
	\IEEEauthorblockN{}
	\IEEEauthorblockA{Ji Wang, \IEEEmembership{Senior Member,~IEEE}, Jiayu Tian, Lijuan Qin, Kunrui Cao, Hongbo Xu, \\ Xingwang Li, \IEEEmembership{Senior Member,~IEEE}, and Tony. Q. S. Quek, \IEEEmembership{Fellow,~IEEE}}
	
	\thanks{Ji Wang, Jiayu Tian, Lijuan Qin and Hongbo Xu are with the Department of Electronics and Information Engineering, College of Physical Science and Technology, Central China Normal University, Wuhan 430079, China (e-mail:jiwang@ccnu.edu.cn;tianjiayu@mails.ccnu.edu.cn;lijuanqincl@mails.cc-nu.edu.cn;xuhb@ccnu.edu.cn).(\emph{Corresponding author:Hongbo Xu})}
	
	\thanks{Kunrui Cao is with School of Information and Communications, National University of Defense Technology, Wuhan, 430035, China (e-mail:krcao@nudt.edu.cn).}
	
	\thanks{Xingwang Li is with the School of Physics and Electronic Information Engineering, Henan Polytechnic University, Jiaozuo 454003, China (e-mail: lixingwang@hpu.edu.cn).}
	
	\thanks{Tony. Q. S. Quek is with the Singapore University of Technology and Design, Singapore 487372, and also with the Yonsei Frontier Lab, Yonsei University, South Korea (e-mail: tonyquek@sutd.edu.sg).}}

\date{}

\begin{document}
	
	\maketitle
	
	\begin{abstract}
		Reconfigurable Intelligent Surfaces (RIS) dynamically control signal propagation to enhance wireless communications. This paper presents a novel framework for rotatable RIS assisted physical-layer multicast systems, aiming to maximize the sum of minimum multicast rates via joint optimization of base station beamforming, RIS phase shifts, and orientation. Unlike unicast or non-rotatable setups, the rotatable RIS adapts orientation to align signals with user groups, improving fairness and rates for weak users. An alternating optimization approach combines convex optimization for beamforming/phase shifts with exhaustive search and particle swarm optimization (PSO) for orientation. Majorization-Minimization-based algorithms solve subproblems iteratively. Simulation results show the framework achieves 24.1\% rate improvement via exhaustive search and 20.0\% via PSO over the non-rotatable RIS baseline, with PSO performance close to the exhaustive search upper bound, highlighting the benefits of physical-layer multicast and orientation optimization.
		
	\end{abstract}
	
	\vspace{-10pt}
	\section{Introduction}
	Reconfigurable Intelligent Surface (RIS) has gained significant attention in wireless communications for their ability to dynamically manipulate signal propagation environments, thereby enhancing system performance \cite{wu2019},\cite{huang2020}. By adjusting the phase shift of passive reflective elements, RIS can effectively increase signal strength and mitigate interference, making it a promising technology for physical-layer multicast systems \cite{zheng2020,Wang2024,Xiao2024}. In particular, RIS-aided systems offer substantial benefits in physical-layer multicast scenarios, where multiple users within the same group share identical data, as opposed to unicast scenarios where each user receives distinct data streams. physical-layer multicast transmission reduces redundant data transmission, improving spectral efficiency, but it also introduces challenges in managing inter-group interference and ensuring fairness among users within each group \cite{shamai2003}.
	
	Conventional RIS architectures with static orientations improve coverage but lack adaptability in dynamic channels. By contrast, our rotatable RIS design unlocks new spatial flexibility through real-time orientation control, dynamically steering reflections toward user clusters\cite{Jiang2025}. This capability is uniquely critical for multicast fairness, where orientation directly impacts group-level Signal to Interference plus Noise Ratio (SINR) balancing \cite{pan2020}. This flexibility is particularly advantageous in physical-layer multicast systems, where the RIS can dynamically adjust its orientation to optimize signal strength for the weakest user in each group, thereby ensuring fairness and maximizing the sum of minimum rates across groups. Compared to a non-rotatable RIS, a rotatable RIS can better mitigate inter-group interference by steering reflections more precisely, leading to significant rate improvements \cite{pan2020}. Furthermore, unlike unicast scenarios where RIS optimization focuses on individual user links, physical-layer multicast systems require balancing the performance across multiple users per group, making the joint optimization of RIS orientation, phase shifts, and beamforming even more critical \cite{Chen2024}.
	
	In this paper, we propose a novel framework to maximize the sum of minimum rates in an RIS assisted physical-layer multicast system. We jointly optimize the base station (BS) beamforming vectors, RIS phase shifts, and RIS orientation angle under power and phase constraints, leveraging the advantages of physical-layer multicast transmission and a rotatable RIS. To solve this non-convex problem, we employ an alternating optimization (AO) strategy \cite{wu2020AO}, combining convex optimization for beamforming and phase shifts with an exhaustive search for RIS orientation. We further develop efficient algorithms based on the Majorization-Minimization (MM) method to optimize beamforming and phase shifts \cite{sun2017}. Using particle swarm optimization (PSO) \cite{kennedy1995} and an exhaustive search optimization algorithm to solve the problem of RIS rotation. Simulation results demonstrate that our approach significantly outperforms traditional non-rotatable RIS setups and unicast systems, achieving higher rates and better fairness among users.

	\vspace{-10pt}
	\section{System Model}	
	
	As illustrated in Fig. 1, we consider a multi-group physical-layer multicast system augmented by a rotatable RIS. The RIS dynamically steers beam from the BS toward user groups to enhance received signal quality \cite{zheng2020}. Assuming that the path is heavily attenuated, signals that experience multiple reflections are ignored \cite{direnzo2019}. Let $M$ denote the number of RIS elements. Each RIS element reflects the beam independently by phase shifting, which together forms a matrix of reflection coefficients ${\bf{E}} = {\rm{diag}}\left( {{{\left[ {{e_1},...,{e_M}} \right]}^{\rm{T}}}} \right){ \in\mathbb{C}^{M \times M}}$ , where ${\left| {{e_m}} \right|^2} = 1$, $\forall m = \{ 1, \ldots ,M\}$. While the actual amplitude of reflection is largely dependent on the phase shift, we assume that the amplitude follows a unit modulus constraint; Amplitude-phase coupling is deferred to future work. ${{\bf{H}}_{{\rm{BI}}}}{ \in\mathbb{C}^{M \times N}}$ and ${{\bf{h}}_{{\rm{IU,}}k}}{ \in\mathbb{C} ^{M \times 1}}$ represent the channel response from BS to RIS and from RIS to user $k$, respectively. 
\vspace{-10pt}
	\subsection{Channel Model}
	\begin{figure}[t]
		\centering
		\includegraphics[width=0.85\linewidth,clip,trim=5 3 3 3]{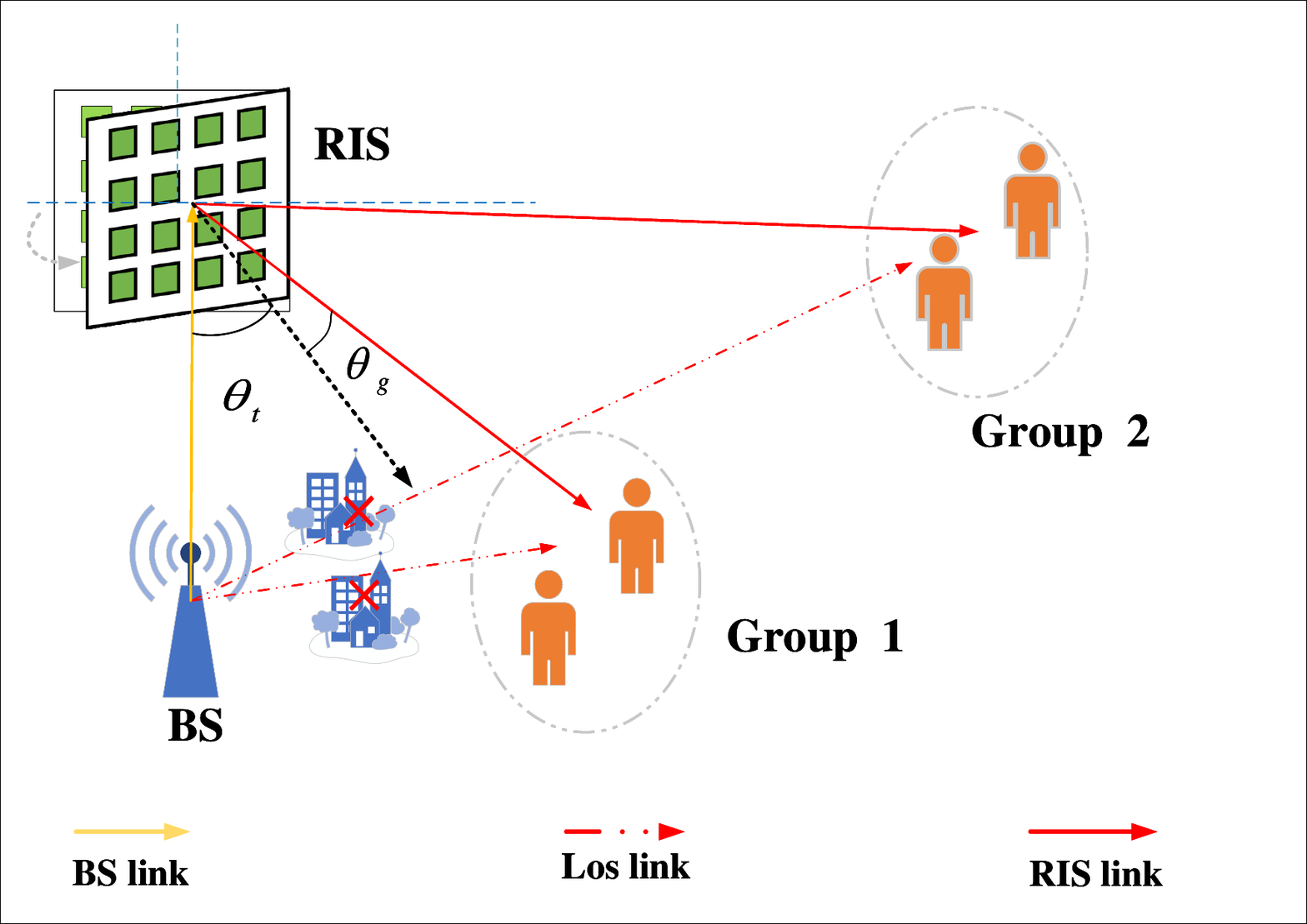}
		\caption{RIS assisted physical-layer multicast system}
		\label{fig:system_model}
	\end{figure}
	
	The channel model incorporates line-of-sight (LoS) and non-line-of-sight (NLoS) components. The BS-RIS channel ${{\rm{H}}_{{\rm{BI}}}}$ is modeled as
	\begin{align}
	{{\bf{H}}_{{\rm{BI}}}} = \sqrt{\text{PL}_{\text{BI}}} \left( \sqrt{\frac{\kappa_{\text{BI}}}{\kappa_{\text{BI}}+1}} {{\bf{H}}_{{\rm{LOS,BI}}}} + \sqrt{\frac{1}{\kappa_{\text{BI}}+1}} {{\bf{H}}_{{\rm{NLOS,BI}}}} \right),\label{eq2}
	\end{align}
	where $\mathrm{PL}_{\mathrm{BI}}$ is the path loss, $\kappa_{\mathrm{BI}}$ is the Rician factor, and ${{\bf{H}}_{{\rm{LOS,BI}}}}$, ${{\bf{H}}_{{\rm{NLOS,BI}}}}$ represent LoS and NLoS components, respectively. Similarly, the RIS-to-user channel ${{\rm{h}}_{{\rm{IU,}}k}}$ for user $k$ is modeled with path loss $\mathrm{PL}_{\mathrm{IU},k}$ and Rician factor $\kappa_{\mathrm{IU},k}$.
	\vspace{-10pt}
	\subsection{Transmission Model}
	Assuming the model is located in a far-field environment for the BS and users relative to the RIS, the distance between the BS (or user $k$) and RIS elements is approximated as equal. The gain of RIS is\cite{Cheng2022}
	\begin{align}
	{\xi _g} = {G_t}{G_g}{\rm{diag}}\left( e \right){\rm{ }} \buildrel \Delta \over = D_m^2{\cos ^q}\left( {{\theta _t}} \right){\cos ^q}\left( {{\theta _g}} \right){\rm{diag}}\left( e \right),\label{eq}
    \end{align}
    where $D_m$ is the maximum RIS directivity, ${G_t}$ is the receiving gain from BS to RIS, $G_g$ is the transmitting gain from RIS to group $g$, and ${\cos ^q}\left( {{\theta _t}} \right)$, $\cos^q(\theta_g)$ model the Lambertian radiation patterns from BS to RIS and RIS to group $g$, respectively \cite{wang2023}. Where ${\theta _t}$ is denoted as the angle of arrival from BS to RIS, and ${\theta _k}$ indicates the angle of departure from the RIS to group $g$. The received signal for user $k$ is
	\begin{align}
	{{\rm{y}}_k} = {\bf{h}}_{{\rm{IU}},k}^{\rm{H}}{\xi _g}{{\bf{H}}_{{\rm{BI}}}}\sum\nolimits_{g = 1}^G {{f_g}{s_g}}  + {n_k},\label{eq6}
    \end{align}
	where $\mathbf{e} \in \mathbb{C}^{M \times 1}$ is the phase shift vector with $|e_m| = 1$, ${f_g} \in \mathbb{C}^{N \times 1}$ is the beamforming vector for group $g$, $s_g$ is the transmitted symbol, and $n_k \sim \mathcal{CN}(0, \sigma_k^2)$ is the noise.
\vspace{-10pt}
	\subsection{Problem Formulation}
	This paper focuses on maximizing the sum of the minimum achievable rates across user groups in a rotatable RIS-aided physical-layer multicast system, ensuring equitable performance within each group \cite{palomar2006}. The achievable rate for user $k$ in group $g$ is defined as
	\begin{align}
	{R_k} = {\log _2}\left( {1 + \frac{{{{\left| {{G_t}{G_g}{\bf{h}}_{{\rm{IU,}}k}^{\rm{H}}{\rm{diag}}\left( e \right){{{\bf{H}}_{{\rm{BI}}}}}{f_g}} \right|}^2}}}{{\sum\nolimits_{i \ne g}^G {{{\left| {{G_t}{G_g}{\bf{h}}_{{\rm{IU,}}k}^{\rm{H}}{\rm{diag}}\left( e \right){{{\bf{H}}_{{\rm{BI}}}}}{f_i}} \right|}^2} + \sigma _k^2} }}} \right),
    \end{align}
    where $\mathbf{f}_g$ is the beamforming vector for group $g$, $e$ is the RIS phase shift vector, and $\sigma_k^2$ is the noise variance. By exploiting the spatial flexibility of a rotatable RIS, we propose a joint optimization of the precoding matrix $\mathbf{F} = [\mathbf{f}_1, \ldots, \mathbf{f}_G]$, phase shift vector $e$, and rotation angle $\delta$ to maximize the group-fairness metric \cite{yu2021}:	 
	\begin{align}
		\max_{{\bf{F}}, e, \delta} \quad&\sum_{g=1}^G \min_{k \in \kappa_g} R_k({\bf{F}}, e, \delta) \tag{5a} \label{eq5a} \\
		\text{s.t.} \quad &{\bf{F}} \in S_F, \, e \in S_e, \, \delta \in \left(-\frac{\pi}{2}, \frac{\pi}{2}\right), \tag{5b} \label{eq5b}
	\end{align}
	where $S_F$ and $S_e$ denote the feasible sets for $\mathbf{F}$ and ${e}$, respectively, and $\delta$ is constrained within the feasible rotation range. This optimization problem, denoted as (4), is non-convex due to the non-differentiable objective function $F(\mathbf{F}, {e}, \delta)$ and the non-convex unit-modulus constraint $S_e$. To address these challenges, we develop efficient algorithms in the subsequent section.
	\vspace{-10pt}
	\section{Proposed Solutions}

	To tackle the non-convex optimization problem (4), we introduce a set of algorithms combining second-order cone programming (SOCP) within the MM framework, PSO, and an exhaustive search approach \cite{wu2020AO,sun2017}. The MM method first approximates the non-convex objective function with a concave surrogate, enabling iterative solutions for subproblems involving different variables. The AO strategy is employed to update the precoding matrix $\mathbf{F}$, phase shift vector ${e}$, and rotation angle $\delta$ alternately, addressing the composite nature of $F(\mathbf{F}, {e}, \delta)$, which involves linear combinations of pointwise minima with non-concave subfunctions $R_k(\mathbf{F}, {e}, \delta)$ \cite{zhou2020}, expressed as
    \[\begin{array}{l}
	{\widetilde R_k}\left( {{\bf{F}},e,\delta \left| {{{\bf{F}}^n},{e^n},{\delta ^n}} \right.} \right)\\
	={\rm{cons}}{{\rm{t}}_k} + 2{\rm{Re}}\left\{ {{a_k}{G_t}{G_g} {e^{\rm{H}}}{{\bf{H}}_k}{f_g}} \right\}-{b_k}\sum\limits_{i = 1}^G {{{\left| {{G_t}{G_g}{e^{\rm{H}}}{{\bf{H}}_k}{f_i}} \right|}^2}} \\
	\le {R_k}\left( {{\bf{F}},e,\delta } \right), \tag{6} \label{eq6}
    \end{array}\]\\
where $\mathbf{H}_k = \operatorname{diag}\left( \mathbf{h}_{\text{IU}}^{\text{H}} \right) \mathbf{H}_{\text{BI}}$, $a_k = \frac{\zeta _k }{\eta_k }$, $b_k =\frac{{{{\left( {{\zeta_k ^{\rm{H}}}} \right)}^2}}}{{\eta_k \left( {{{\left( {{\zeta_k ^{\rm{H}}}} \right)}^2} + \eta_k } \right)}}$, ${\rm{cons}}{{\rm{t}}_k} = {R_k}\left( {{{\bf{F}}^n},{e^n},{\delta ^n}} \right) - {b_k}\sigma _k^2 - {b_k}\eta_k$, ${\zeta _k} = G_t^nG_g^n{\left( {f_g^n} \right)^{\rm{H}}}{\bf{H}}_k^{\rm{H}}{e^n}$, ${\eta _k} = \sum\limits_{i \ne g}^G {{{\left| {G_t^nG_g^n{{\left( {{e^n}} \right)}^{\rm{H}}}{{\bf{H}}_k}f_i^n} \right|}^2}}  + \sigma _k^2$, at fixed point $\left\{ {{{\bf{F}}^n},{e^n},{\delta ^n}} \right\}$.	
Proof refer to \cite{zhou2020}. Problem (5) can be rewritten as
	\begin{align}
		\mathop {\max }\limits_{{\bf{F}},e,\delta } \quad &\widetilde F\left( {{\bf{F}},e,\delta \mid {{\bf{F}}^n},{e^n},{\delta ^n}} \right) \tag{7a} \label{eq7a} \\
		\text{s.t.} \quad &{\bf{F}} \in S_F, \, e \in S_e, \, \delta \in \left(-\frac{\pi}{2}, \frac{\pi}{2}\right). \tag{7b} \label{eq7b}
	\end{align}
	where \[\mathop {\max }\limits_{{\bf{F}},e,\delta } \widetilde F\left( {{\bf{F}},e,\delta \mid {{\bf{F}}^n},{e^n},{\delta ^n}} \right) = \sum\limits_{g = 1}^G {\mathop {\min }\limits_{k \in {K_g}} } {\widetilde R_k}\left( {{\bf{F}},e,\delta \mid {{\bf{F}}^n},{e^n},{\delta ^n}} \right).\]
	
	We note that ${{{\widetilde R}_k}\left( {{\bf{F}},e,\delta \left| {{{\bf{F}}^n},{e^n},{\delta ^n}} \right.} \right)}$ is biconcave of ${\bf{F}}$ and $e$, since ${\widetilde R_k}\left( {{\bf{F}}\left| {{{\bf{F}}^n}} \right.} \right) = {\widetilde R_k}\left( {{\bf{F}},e,\delta \left| {{{\bf{F}}^n},{e^n},{\delta ^n}} \right.} \right)$ with given $e$ is concave of ${\bf{F}}$ and ${\widetilde R_k}\left( {e\left| {{e^n}} \right.} \right) = {\widetilde R_k}\left( {{\bf{F}},e,\delta \left| {{{\bf{F}}^n},{e^n},{\delta ^n}} \right.} \right)$ with given ${\bf{F}}$ is concave of $e$. This biconvex problem enables us to use the AO method to alternately update ${\bf{F}}$ and $e$. The rotation angle $\delta $ and the RIS phase shift vector $e$ are tightly coupled, and the objective function’s non-convex nature makes it challenging for conventional algorithms to find the optimal solution without exhaustively searching all possible angles, which is computationally intensive. To address this issue, we employed the PSO algorithm to efficiently optimize the rotation angle $\delta $.
	\vspace{-15pt}
	\subsection{Beamforming Matrix ${\bf{F}}$ Optimization}	
	This section focuses on optimizing the precoding matrix $\mathbf{F}$ while keeping the RIS phase shift vector ${e}$ and rotation angle $\delta$ fixed. The surrogate rate function $\tilde{R}_k(\mathbf{F} \mid \mathbf{F}^n)$, derived in \cite{sun2017}, is expressed as a quadratic form of $\mathbf{F}$:
	\[
	\begin{array}{l}{{R_k}\left( {{\bf{F}}\left| {{{\bf{F}}^n}} \right.} \right) = {\rm{cons}}{{\rm{t}}_k} + 2{\rm{Re}}\left\{ {{\rm{Tr}}\left[ {{\rm{C}}_k^{\rm{H}}{\bf{F}}} \right]} \right\} - {\rm{Tr}}\left[ {{{\bf{F}}^{\rm{H}}}{{\rm{B}}_k}{\bf{F}}} \right],\tag{8} \label{eq8}}
	\end{array}
	\]
	where ${{\rm{B}}_k} = {b_k}G_t^nG_g^n{\bf{H}}_k^{\rm{H}}e{e^{\rm{H}}}{{\bf{H}}_k}$, ${\rm{C}}_k^{\rm{H}} = {a_k}G_t^nG_g^n{{\rm{t}}_g}{e^{\rm{H}}}{{\bf{H}}_k}$, and ${{\rm{t}}_g}{ \in \mathbb{R} ^{G \times 1}}$ is a selection vector with the $g$-th element set to one and others to zero. According to (8), The optimization subproblem is	
	\begin{align}
		\mathop {\max }\limits_{\bf{F}}\quad&\sum\limits_{g = 1}^G {\mathop {\min }\limits_{k \in {K_g}} \left\{ {{\rm{cons}}{{\rm{t}}_k} + 2{\rm{Re}}\left\{ {{\rm{Tr}}\left[ {{\rm{C}}_k^{\rm{H}}{\bf{F}}} \right]} \right\} - {\rm{Tr}}\left[ {{{\bf{F}}^{\rm{H}}}{{\rm{B}}_k}{\bf{F}}} \right]} \right\}}\tag{9a} \label{eq9a} \\
		{\rm{s}}.{\rm{t}}.\;\quad&{\bf{F}} \in {S_F}.{\rm{ }}\tag{9b} \label{eq9b}
	\end{align}
	To address the minimum operation, auxiliary variables $\gamma = [\gamma_1, \ldots, \gamma_G]^{\mathrm{T}}$ are introduced, transforming the problem (9) into a SOCP form:
	\begin{align}
		\mathop {\max }\limits_{\bf{F}} {\rm{ }}\quad&\sum\limits_{g = 1}^G {{\gamma _g}}  \tag{10a} \label{eq10a} \\
		\text{s.t.} \quad& {\bf{F}} \in S_F, \tag{10b} \label{eq10b} \\
		\quad& \text{const}_k + 2 \operatorname{Re} \left\{ \operatorname{Tr}\left[ {\bf{C}}_k^{\mathrm{H}} {\bf{F}} \right] \right\} - \operatorname{Tr}\left[ {\bf{F}}^{\mathrm{H}} {\bf{B}}_k {\bf{F}} \right] \ge \gamma_g. \tag{10c} \label{eq10c}
	\end{align}
	
	This SOCP problem (10) can be effectively solved with the convex optimization tools such as MOSEK, ensuring global optimality for the subproblem \cite{sun2017}.
\vspace{-16pt}
	\subsection{Phase Shift Vector e Optimization}
	
\begin{algorithm}[t!]
\caption{Second-Order Cone Programming for (7)}
\label{alg:Framework}
\begin{algorithmic}[1]
	\STATE Initialize \(\mathbf{F}^0\) and \({e}^0\), set \(n = 0\)
	\REPEAT
	\STATE Compute \({e}^{n+1}\) by solving the optimization problem with \(\mathbf{F}^n\) fixed
	\STATE Compute \(\mathbf{F}^{n+1}\) by solving the optimization problem with \({e}^{n+1}\) fixed
	\STATE \(n \rightarrow n + 1\)
	\UNTIL \(\mathcal{F}(\mathbf{F}, {e}, \delta)\) convergence or maximum iterations reached
	\RETURN \(\mathbf{F}^{n+1}\) and \({e}^{n+1}\)
\end{algorithmic}
\end{algorithm}
\vspace{-5pt}
		This section focuses on optimizing the RIS phase shift vector ${e}$ while keeping the precoding matrix $\mathbf{F}$ and rotation angle $\delta$ fixed. The surrogate rate function $\tilde{R}_k({e} \mid {e}^n)$ can be rewritten as
	\[
	{\widetilde R_k}\left( {e\left| {{e^n}} \right.} \right) = {\rm{cons}}{{\rm{t}}_k} + 2{\mathop{\rm Re}\nolimits} \left\{ {{\rm{a}}_k^{\rm{H}}e} \right\} - {e^{\rm{H}}}{{\rm{A}}_k}e, \tag{11} \label{eq11}
	\]
	where ${{\rm{A}}_k} = {b_k}G_t^nG_k^n{{\bf{H}}_k}\sum\nolimits_{i = 1}^G {{f_i}} f_i^{\rm{H}}{\bf{H}}_k^{\rm{H}}$ and ${{\rm{a}}_k} = {a_k}G_t^nG_k^n{{\bf{H}}_k}{f_g}$. Upon replacing the objective function of problem (7) by (11), the problem rewritten as
	\begin{align}
	\max_e \quad&\sum\nolimits_{g=1}^G \min_{k \in K_g} \left\{ \text{const}_k + 2 \operatorname{Re} \left\{ {\bf{a}}_k^{\mathrm{H}} e \right\} - e^{\mathrm{H}} {\bf{A}}_k e \right\} \tag{12a} \label{eq12a} \\
	\text{s.t.} \quad& e \in S_e. \tag{12b} \label{eq12b}
	\end{align}
	
	To address the pointwise minimum, auxiliary variables $\kappa = [\kappa_1, \ldots, \kappa_G]^{\mathrm{T}}$ are introduced, as follows
	\begin{align}
		\max_{e, \kappa} \quad&\sum_{g=1}^G \kappa_g \tag{13a} \label{eq13a} \\
	    {{\rm{ s}}.{\rm{t}}.} \quad&{e \in {S_e}}, \tag{13b} \label{eq13b} \\
		\quad&\text{const}_k + 2 \operatorname{Re} \left\{ {\bf{a}}_k^{\mathrm{H}} e \right\} - e^{\mathrm{H}} {\bf{A}}_k e \ge \kappa_g. \tag{13c} \label{eq13c} 
	\end{align}
	
    The unit-modulus set $S_e$ is non-convex, which makes the optimization problem still non-convex. To solve this problem, we relax $S_e$ as a convex set:
	\[{S_{e - relax}} = \left\{ {{e^{\rm{H}}}{\rm{diag}}\left( {{i_m}} \right)e \le 1,{\rm{ }}\forall  = 1{\rm{ }},{\rm{ }}{\rm{. }}{\rm{. }}{\rm{. , }}M} \right\},\]
	where ${i}_m \in \mathbb{R}^{M \times 1}$ is a selection vector with the $m$-th element equal to one and others zero. The relaxed problem becomes:
	\begin{align}
		\max_{e, \kappa} \quad&\sum_{g=1}^G \kappa_g \tag{14a} \label{eq14a} \\
		\text{s.t.} \quad &e \in S_{e - \text{relax}}, \tag{14b} \label{eq14b} \\
		\quad&\text{const}_k + 2 \operatorname{Re} \left\{ {\bf{a}}_k^{\mathrm{H}} e \right\} - e^{\mathrm{H}} {\bf{A}}_k e \ge \kappa_g. \tag{14c} \label{eq14c} 
	\end{align}

	Problem (14) is an SOCP problem.  Let $\mathbf{e}^n$ denote the solution to the relaxed problem. The phase is then projected to satisfy the unit-modulus constraint:
	\begin{align}
		e^n = \arg\max_{e} \quad&\sum_{g=1}^G \kappa_g \tag{15a} \\
		\text{s.t.} \quad& e \in S_{e-\text{relax}}, \tag{15b} \\
		\quad& \text{const}_k + 2 \text{Re} \left\{ a_k^{\text{H}} e \right\} - e^{\text{H}} A_k e \geq \kappa_g. \tag{15c} 
	\end{align}
	Specifically, the updated phase shift vector is computed as
	\[{e^{n + 1}} = \exp \left\{ {j\angle \left( {\frac{{{e^n}}}{{{{\left[ {{e^n}} \right]}_M}}}} \right)} \right\}, \tag{16} \label{eq16}\]	
	and symbol $[{e^n}]_m$ represents the $m$-th element of $\mathbf{e}^n$, and the operations $\exp\{\cdot\}$ and $\angle(\cdot)$ are performed element-wise, ensuring each element of ${e}^{n+1}$ satisfies $|e_m| = 1$.
\subsection{RIS Orientation Optimization}

The optimization of the RIS orientation, denoted by the rotation angle \(\delta\), is pivotal for enhancing the performance of the physical-layer multicast system. This section focuses on optimizing \(\delta\) while keeping the precoding matrix \(\mathbf{F}\) and the RIS phase shift vector \(\mathbf{e}\) fixed. The goal is to maximize the sum of the minimum rates across user groups, as defined in the original optimization problem. The non-convex nature of the objective function and the strong coupling between \(\delta\) and \(\mathbf{e}\) pose significant challenges for traditional optimization methods. To overcome these, we propose two approaches: a grid-based search algorithm and an adaptive swarm intelligence algorithm, both designed to efficiently identify a near-optimal \(\delta\) within the feasible range \(\delta \in \left(-\frac{\pi}{2}, \frac{\pi}{2}\right)\).
\begin{algorithm}[t!]
	\caption{Exhaustive Search Algorithm for (17)}
	\begin{algorithmic}[1]
		\STATE Initialize \(\delta \to -\frac{\pi}{2}\), number of discrete points \(D\)
		\STATE Define candidate angles as \(\left\{-\frac{\pi}{2}, -\frac{\pi}{2} + \frac{\pi}{D}, \ldots, \frac{\pi}{2}\right\}\)
		\STATE \textbf{repeat}
		\FOR {each \(\delta\) in candidate set}
		\STATE Compute \(\tilde {F}(\mathbf{F}, {e}, \delta | \mathbf{F}^n, {e}^n, \delta^n)\)
		\ENDFOR
		\STATE Select \(\delta\) that maximizes \(\tilde F\)
		\STATE \textbf{Until} convergence or maximum iterations reached
	\end{algorithmic}
\end{algorithm}

1) Exhaustive Search Optimization Algorithm

The exhaustive search optimization algorithm systematically evaluates the objective function over a discretized set of rotation angles to determine the optimal \(\delta\). The feasible range \(\left(-\frac{\pi}{2}, \frac{\pi}{2}\right)\) is partitioned into \(D\) equidistant points, each representing a candidate angle \(\delta\). For each candidate, the surrogate objective function \(\bar{F}(\mathbf{F}, {e}, \delta | \mathbf{F}^n, {e}^n, \delta^n)\), which approximates the sum of minimum rates, is computed. The angle that maximizes this function is selected as the solution for the current iteration. The accuracy of this method improves with a larger \(D\), but computational complexity increases linearly with the number of discrete points, making it resource-intensive for fine steps.

The subproblem for optimizing \(\delta\) is formulated as
\begin{align}
	\max_{\delta} \quad&\sum_{g=1}^G \min_{k \in K_g} \left[ \text{const}_k + 2 \operatorname{Re} \{ G_t G_g \chi_k \} - G_t^2 G_g^2 \varphi_k \right] \label{eq17a} \tag{17a} \\
	\text{s.t.} \quad&\delta \in \left( -\frac{\pi}{2}, \frac{\pi}{2} \right), \label{eq17b} \tag{17b}
\end{align}
where \(\chi_k \equiv a_k {e}^{\mathrm{H}} \mathbf{H}_k \mathbf{f}_g\) and \(\varphi_k \equiv b_k \sum_{i=1}^G |{e}^{\mathrm{H}} \mathbf{H}_k \mathbf{f}_i|^2\).

2) Particle Swarm Optimization Algorithm

To achieve a balance between solution quality and computational efficiency, we propose an adaptive swarm intelligence algorithm, termed PSO. This algorithm initializes a population of \(Y\) particles, each representing a candidate rotation angle \(\delta\). Each particle has a position \(x_m\) and a velocity \(v_m\), which guide its trajectory through the solution space. The algorithm iteratively updates positions and velocities based on individual and global best solutions, promoting efficient convergence to a near-optimal \(\delta\).

The subproblem for optimizing \(\delta\) is same as (17), The velocity update for the \(m\)-th particle at iteration \(t\) is
\begin{align}
	v_m^t = w^t v_m^{t-1} + c_1 r_1 (p_m^{t-1} - x_m^{t-1}) + c_2 r_2 (g^{t-1} - x_m^{t-1}),\label{eq18} \tag{18}
\end{align}
where \(w^t\) is the inertia weight, \(c_1\) and \(c_2\) are learning factors, \(r_1, r_2 \in [0, 1]\) are random numbers, \(p_m^{t-1}\) is the individual best position, and \(g^{t-1}\) is the global best position. The position is updated as
\begin{align}
	x_m^t = x_m^{t-1} + v_m^t. \label{eq19} \tag{19}
\end{align}

The inertia weight adapts dynamically:
\begin{align}
	w^t = w_{\text{max}} - \left(w_{\text{max}} - w_{\text{min}}\right) \frac{t}{t_{\text{max}}}. \label{eq20} \tag{20}
\end{align}

\begin{algorithm}[t]
	\caption{Particle Swarm Optimization for (17)}
	\begin{algorithmic}[1]
		\STATE Initialize \(Y\) particles with random positions \(x_m^0 \in \left(-\frac{\pi}{2}, \frac{\pi}{2}\right)\), velocities \(v_m^0\), individual best \(p_m^0 = x_m^0\), and global best \(g^0\)
		\FOR {\(t = 1\) to \(t_{\text{max}}\)}
		\STATE Update velocities using velocity update equation
		\STATE Update positions, ensuring \(x_m^t \in \left(-\frac{\pi}{2}, \frac{\pi}{2}\right)\)
		\STATE Evaluate \(\bar{F}\) at each new position
		\STATE Update \(p_m^t\) if new position improves objective
		\STATE Update \(g^t\) if any position improves global best
		\ENDFOR
		\STATE Output \(g^{t_{\text{max}}}\) as optimized \(\delta\)
	\end{algorithmic}
\end{algorithm}

%
\vspace{-8pt}	
	\section{Simulation Results}
	\label{sec:simulation}
We evaluate the performance of the proposed rotatable RIS assisted physical-layer multicast framework through numerical experiments, averaging outcomes over 100 independent trials.	As illustrated in Fig. 1, the system configuration places the BS at coordinates (100 m, 0 m, 0 m) with a uniform linear array (ULA) of $N$ antennas. The RIS, located at (0 m, 0 m, 0 m), employs a uniform planar array (UPA) with $M$ reflecting elements. A total of $K=4$ users, organized into $G=2$ groups, are randomly distributed within a 100 m × 100 m area. Key parameters include noise power $\sigma_k^2 = -164$ dBm, radiation pattern exponent $q=2$, and maximum directivity $D_m=6$.

Three methods are compared, including optimizing RIS phase shifts \(\mathbf{e}\) and BS precoding \(\mathbf{F}\) with fixed RIS orientation; optimizing \(\mathbf{e}\), \(\mathbf{F}\), and RIS rotation angle \(\delta\) using PSO, and optimizing \(\mathbf{e}\), \(\mathbf{F}\), and \(\delta\) as performance upper bounds by exhaustive search. The performance metric is the average sum of minimum rates across groups, \(\sum_{g=1}^G \min_{k \in K_g} R_k\).

In Fig. 2, the convergence trend of each algorithm in terms of the number of iterations is studied. The ${P_{\max }}$ is $20{\rm{dBm}}$ and the step size is $\pi /1440$. Fixed orientation converges to \(5.8 \, \text{bps/Hz}\) in 20--30 iterations. PSO achieves\(\ 6.9 \, \text{bps/Hz}\) in 15--25 iterations. Exhaustive search reaches\(\ 7.2 \, \text{bps/Hz}\) in 10--20 iterations, improving by 4.3\% over \(\pi/1440\) step size and 24.1\% over fixed orientation. Fig.3 shows the relationship between the total rate and the transmit power. Sum rates increase with \(P_{\max}\). PSO and exhaustive search outperform fixed orientation, with exhaustive search yielding the highest rates due to precise RIS alignment.

The results confirm the effectiveness of the proposed framework, with exhaustive search obtaining the highest rate, and PSO providing a practical, low-complexity alternative. Then, system performance can be further improved by adding a more granular exhaustive search step size, which provides the best sum rate at the cost of higher complexity, and PSO-based approaches that provide a real trade-off. The proposed framework effectively solves the non-convexity of the joint optimization problem, and achieves robust performance through improved angular resolution.
 
	\begin{figure}[t]
		\centering
		\includegraphics[width=0.9\linewidth,clip,trim=0 0 0 0]{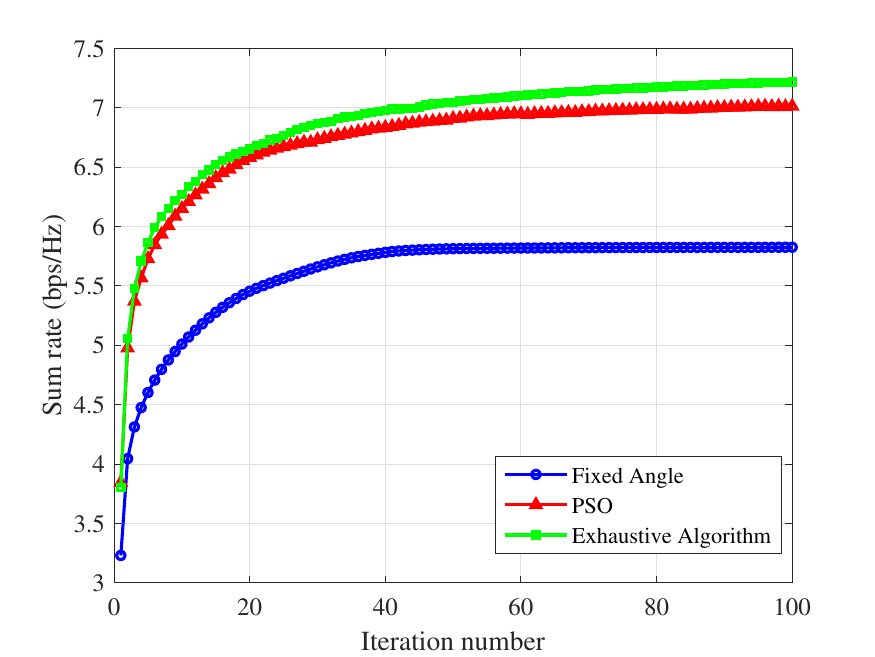}
		\caption{Sum rate versus iteration number}
		\label{fig.3}
	\end{figure}
  \vspace{-5pt}
	\begin{figure}[t]
		\centering
		\includegraphics[width=0.9\linewidth,clip,trim=0 0 0 0]{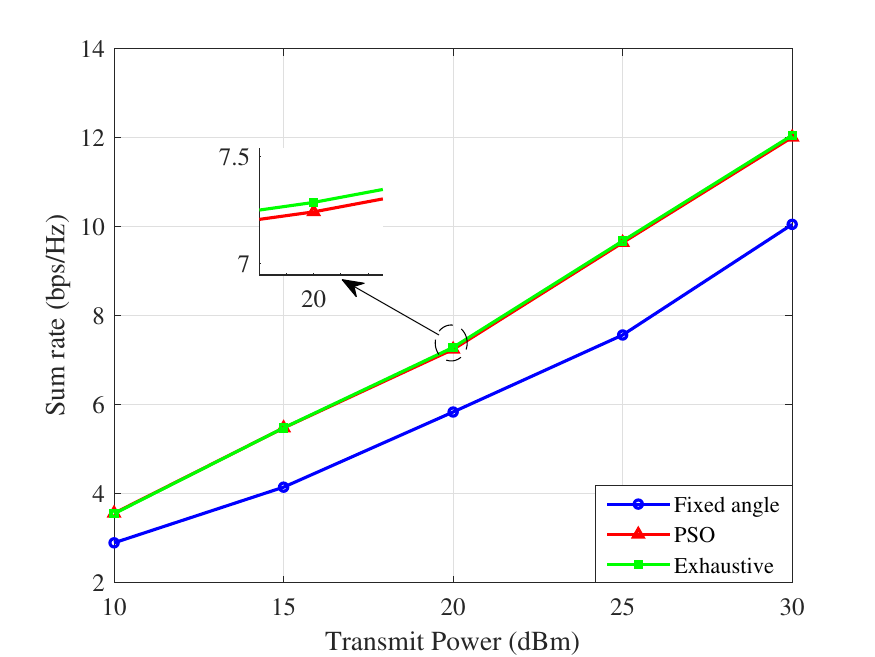}
		\caption{Sum rate versus transmit power}
		\label{fig.4}
	\end{figure}  
	\vspace{-10pt}	
	\section{Conclusion}
	This paper proposes a joint optimization framework for rotatable RIS-aided multi-user MISO physical layer multicast systems. The framework focuses on jointly designing BS beamforming, RIS phase shifts, and RIS orientation. By leveraging a rotatable RIS, the framework dynamically aligns reflected signals with user groups, addressing inter-group interference and fairness challenges. An alternating optimization strategy combining the MM method and PSO is developed to handle the non-convex nature of the problem. Simulation results validate that the rotatable RIS significantly enhances system performance compared to fixed-orientation RIS setups, with the PSO algorithm achieving near-optimal performance close to the upper bound of exhaustive search. Future work will explore low-complexity RIS orientation optimization algorithms and incorporate practical RIS amplitude models to facilitate real-world deployment.

\vspace{-10pt}
\bibliography{ref}
\bibliographystyle{IEEEtran}
	
\end{document}